\documentclass[12pt]{article}
\usepackage{color}
\usepackage{epsfig}
\usepackage{amsmath}
\usepackage{hhline}
\usepackage{amsfonts,amssymb}
\usepackage{times}

\graphicspath{{./eps/}}
\DeclareGraphicsExtensions{.eps,.eps.gz,.ps,.ps.gz}
\newlength{\dinwidth}
\newlength{\dinmargin}
\newcommand{\GeV} {\mathrm{GeV}}
\newcommand{\TeV} {\mathrm{TeV}}
\newcommand{\fb}  {\mathrm{fb}}

\newcommand{\mrad} {\mathrm{mrad}}

\newcommand{\cL } {{\cal L}}
\newcommand{\cP } {{\cal P}}
\newcommand{\cB } {{\cal B}}

\def\pythia  {{\sc Pythia}}

\def\ee    {e^+e^-}
\def\ti    {\tilde}

\def\stau  {{\ti\tau}}

\def\staup {{\ti\tau}^+}

\def\cx    {\ti {\chi}}

\def\cp    {\ti {\chi}^+}
\def\cm    {\ti {\chi}^-}

\def\nt    {\ti {\chi}^0}

\def\smur  {{\ti\mu}_R}

\def\snm   {{\ti\nu}_\mu}

\def\ser   {{\ti e}_R}

\def\sne  {{\ti\nu}_e}

\def\snt      {{\ti\nu}_\tau}

\def \Eslash {E \kern-.6em\slash }
\def \Mslash {M \kern-.5em\slash }

\newcommand{\nn}{\nonumber}
\newcommand{\beq}{\begin{equation}}
\newcommand{\eeq}{\end{equation}}
\newcommand{\bea}{\begin{eqnarray}}
\newcommand{\eea}{\end{eqnarray}}

\newcommand{\fig}[1]{figure~\ref{#1}}
\newcommand{\tab}[1]{table~\ref{#1}}

\begin{document}

\title{  
  \begin{flushright} {\normalsize
      DESY 05--150 }
  \end{flushright}
  \vspace{1cm}  
  \LARGE\bfseries
  Determining the Stau Trilinear Coupling {\boldmath $A_\tau$} \\[2mm]
  in Supersymmetric Higgs Decays }
\author{
  S.Y.~Choi$^1$, H.--U.~Martyn$^2$ and P.M.~Zerwas$^3$
  \\[.5ex]
  { \itshape \small $ ^1$
    Department of Physics, Chonbuk National University, Chonju, Korea} \\
  { \itshape \small
     $ ^2$ I. Physikalisches Institut, RWTH Aachen, Germany} \\
  { \itshape \small $ ^3$
  Deutsches Elektronensynchrotron DESY, Hamburg, Germany}
}
\date { }

\maketitle

\thispagestyle{empty}

\begin{quote}   \small
The measurement of the trilinear couplings $A$ in the part of
the Lagrangian which breaks supersymmetry softly will be a
difficult experimental task. In this report
the heavy Higgs decays $H,A\to \stau_1\stau_2$ to stau pairs
are investigated for measuring the stau trilinear coupling $A_\tau$.
Based on detailed simulations of signal and backgrounds
for a specific reference point in future high luminosity
$e^+e^-$ linear collider experiments,
it is concluded that
the parameter $A_\tau$ can be determined with a precision at the 10\% level  in
the region of moderate to large $\tan\beta$.
\end{quote}

\section{Introduction}
  \label{intro}

The couplings between fermionic matter fields and Higgs fields
differ from those of the scalar matter fields once supersymmetry
is broken, see {\it e.g.} Ref.\cite{Djou}.
In theories based on soft supersymmetry breaking the
scalar-Higgs Yukawa couplings are modified multiplicatively
by the $A$ parameters which, in parallel
to the fermion-Higgs Yukawa couplings, are inter-generational matrices. 
In accordance with bounds on flavor-changing couplings
the $A$ parameters are generally assumed to be
diagonal and three parameters, $A_t$, $A_b$ and $A_\tau$,
are introduced for the third generation.

By definition, the $A$ parameters come with the Yukawa couplings
which are of the size of the fermion masses.
Therefore they cannot
be measured in general directly except for the third generation. Since they
couple Higgs fields with scalar $L$-fields and $R$-fields, they
become effective in two ways: (i) They contribute to the
off-diagonal elements in the scalar mass matrices, and to the
mixing of $L$- and $R$-states; and (ii) They give rise
to mixed scalar $L$ and $R$ decay final states of the heavy scalar
and pseudoscalar Higgs bosons.

In the scalar stop sector the off-diagonal mass matrix element is
given by $m_t (A_t - \mu \cot\beta)$. For moderate to large
$\tan\beta$ the second term is suppressed and $A_t$ can be
determined quite accurately by measuring the mixing effects in the
stop mass spectrum and the stop mixing angle in $e^+e^-$
annihilation to stop pairs \cite{bartl}. In heavy Higgs decays, on
the other hand, $A_t$ is shielded by the potentially much larger term
$\mu \tan\beta$ and Higgs decays to stop pairs, if kinematically
allowed at all, are less suited for measuring the stop trilinear
parameter.

The situation is reversed in the down sector, {\em i.e.} for staus.
While in the stau system $A_\tau$ is shielded by the term $\mu
\tan\beta$ in the mass matrix~\cite{boos}, the $A_\tau$ parameter
is enhanced by the coefficient $\tan\beta$
in the couplings of the heavy scalar and pseudoscalar Higgs bosons
to mixed pairs of $\stau_L$ and $\stau_R$ fields.
Heavy Higgs decays are therefore
promising channels for measuring the stau trilinear parameter $A_\tau$.

The expressions for the partial decay widths become especially
transparent in the limit where (i) the heavy Higgs boson masses
are large [decoupling limit], (ii) $\tan\beta$ is large, and (iii)
the $LR$ mixing is small. In this limit the decay widths of the
scalar and pseudoscalar Higgs bosons to mixed pairs
$\stau_1\stau_2 \equiv \stau_1^+\stau_2^- + \stau_1^-\stau_2^+$
are given by
\begin{eqnarray}
  \Gamma(H,A \to \stau_1\stau_2) & \simeq &
  \frac{G_F m_\tau^2}{4\sqrt{2} \pi} \, \lambda^{1/2}  \,
  \frac{(A_\tau \tan\beta + \mu)^2}{m_{H,A}} \ ,
  \label{eq:gammat12}
\end{eqnarray}
where $\lambda$ accounts for the pase space suppression in the
usual form. The couplings of the scalar Higgs boson $H$ to diagonal
pairs of $L$- and $R$-fields are suppressed by coefficients $m_\tau /
A_\tau$ and $m_Z / (A_\tau \tan\beta)$ which both are small in
the limit we are considering. The coupling of the pseudoscalar
Higgs boson $A$ to diagonal pairs vanishes in CP-invariant
theories.

Using the partial widths for Higgs decays to tau pairs,
\begin{eqnarray}
  \Gamma(H,A \to \tau \tau) & \simeq &
  \frac{G_F m_\tau^2}{4\sqrt{2} \pi} \,
   m_{H,A}\, \tan^2\beta  \ ,  \label{eq:gammatt}
\end{eqnarray}
the decay widths to stau pairs may be
normalized by the decays to tau pairs:
\begin{eqnarray}
  \frac{\Gamma(H,A \to \stau_1\stau_2)}{\Gamma(H,A \to \tau\tau)}
  & \simeq & 
  \lambda^{1/2} \,
      \frac{(A_\tau + \mu\,\cot\beta)^2}{m^2_{H,A} } \ .
      \label{eq:ratiostautau}
\end{eqnarray}
If the normalization is chosen alternatively by the dominating $b \bar{b}$
final states, the ratio of the widths is reduced by a coefficient
$m^2_\tau / 3 m^2_b$:
\begin{eqnarray}
  \frac{\Gamma(H,A \to \stau_1\stau_2)}{\Gamma(H,A \to b \bar{b})}
  & \simeq & 
  \lambda^{1/2} \, \frac{m^2_\tau}{3 m^2_b} \,
      \frac{(A_\tau + \mu\,\cot\beta)^2}{m^2_{H,A} } \ .
      \label{eq:ratiobb}
\end{eqnarray}
In any case, for moderate to large
$\tan\beta$ and $A_\tau$ of the same order as $\mu$, the size of
the branching ratio of the heavy Higgs bosons to mixed $LR$ stau pairs
is essentially set by $A_\tau^2$. Thus, for sufficiently large
$A_\tau$ the measurement of these branching ratios provides a valuable
instrument for measuring $A_\tau$.

\section{Properties of the Higgs system}

The qualitative arguments presented above appear strong enough
to perform a quantitative analysis in order to prove this method
to be useful  for measuring $A_\tau$ in practice.
For this purpose we adopt the mSUGRA
reference point SPS1a$'$ defined for the
{\em SPA Project}~\cite{spa}.
It is closely related to the standard reference point SPS1a,
yet with a cold dark matter density in accordance with
the WMAP measurement.

The mSUGRA parameters are defined as
$M_0=70\;\GeV$, $ M_{1/2} = 250 \;\GeV$,  $A_0 = -300 \;\GeV$,
$\tan\beta = 10$ and $\mbox{sign}~\mu = +$.
Extrapolation to the electroweak scale generates the Lagrangian parameters
$A_\tau = -445~\GeV$ and $\mu = 403~\GeV$, thus $|A_\tau| \gg \mu\cot\beta$
holds indeed.
The masses and branching ratios of the supersymmetric particles relevant to
the present analysis are summarized in \tab{tab:decaymodes}.

\begin{table}[htb] \centering 
  $\begin{array}{|l|c|cc|cc|}
   \hline & & & & & \\[-2.ex]
   \mbox{Particle} & \ \mbox{Mass}\,[\GeV] \ &
   \ \ \ \mbox{Decay} \ \ \ & ~~~~~ \mathcal{B} ~~~~~ &
   \ \ \ \mbox{Decay} \ \ \ & ~~~~~ \mathcal{B} ~~~~~
   \\[.5ex] \hline \hline
   H^0      &  431.1 & \tau^-   \tau^+        & 0.075 & \nt_1 \nt_1  & 0.011 \\
            &        & b        \bar b        & 0.683 & \nt_1 \nt_2  & 0.040 \\
            &        & t        \bar t        & 0.053 & \nt_2 \nt_2  & 0.023 \\
            &        & \stau^-_1 \stau^+_1    & 0.014 & \cp_1 \cm_1  & 0.056 \\
            &       & \stau^\mp_1 \stau^\pm_2 & 0.031 &              & \\
            &        & \stau^-_2 \stau^+_2    & 0.003 &              & \\
   \hline
   A^0      &  431.0 & \tau^-   \tau^+     & 0.055 & \nt_1 \nt_1  & 0.011 \\
            &        & b        \bar b     & 0.505 & \nt_1 \nt_2  & 0.055 \\
            &        & t        \bar t     & 0.103 & \nt_2 \nt_2  & 0.063 \\
           &     & \stau^\mp_1 \stau^\pm_2 & 0.035 & \cp_1 \cm_1  & 0.170
   \\[.5ex] \hline \hline
   \nt_1    &   97.8 &                      &       &                & \\
   \hline
   \nt_2    &  184.4 & \stau_1^\pm\tau^\mp  & 0.564 & \snt \nu_\tau  & 0.155 \\
            &        & \ser^\pm e^\mp       & 0.024 & \sne \nu_e     & 0.115 \\
            &        & \smur^\pm \mu^\mp    & 0.026 & \snm \nu_\mu   & 0.115 \\
   \hline 
   \cp_1    &  184.2 & \staup_1 \nu_\tau  & 0.519 & \snt \tau^+  & 0.189 \\
            &        &                    &    & \sne e^+    & 0.138 \\
            &        &                    &    & \snm \mu^+  & 0.138
    \\[.5ex] \hline \hline
   \stau_1  &  107.4 & \nt_1    \tau^-  & 1.000 &                &\\
   \hline
   \stau_2  &  195.3 & \nt_1    \tau^-  & 0.869 & \cm_1 \nu_\tau & 0.086 \\
            &        & \nt_2    \tau^-  & 0.046 &                &\\
   \hline
   \snt     &  170.7 & \nt_1    \nu_\tau& 1.000 &                &\\
   \hline
   \end{array}$
   \caption{ 
     Masses and branching ratios of heavy Higgs bosons, light gauginos and
     third generation sleptons
     in the SPS1a$'$ scenario~\cite{spa}.
     The Higgs decays are calculated with the program
     {\sc FeynHiggs~2.2.10}~\cite{feynhiggs}}
   \label{tab:decaymodes}
 \end{table}

At a linear collider with an energy $\sqrt{s}$ of about 1 TeV
heavy Higgs boson production $\ee\to HA$, see~\cite{Ohm},
will clearly be kinematically accessible for this reference point~\cite{tdr}.
The measurement of their decay modes, however, will
confront several problems:
\begin{itemize}
  \item[--]
    Due to their mass degeneracy the decays of $H$ and $A$ cannot be
    resolved experimentally. Thus one can only determine the branching ratios
    for the sum of both Higgs bosons.
  \item[--]
    The energy spectra of the final $\tau$ decay products reflect only
    weakly the energy of the primary particles,
    which is gradually softened during cascade decays involving massive
    invisible particles like neutralinos or sneutrinos.
    It is therefore extremely difficult to discriminate $\stau_1$ from
    $\stau_2$ decays.
    Instead, only the sum of all $\stau_i\stau_j$ decay modes
    will be determined.
  \item[--]
    As a consequence of the (moderately) large value $\tan\beta = 10$ the
    neutralino $\nt_2$ and chargino $\cx^\pm_1$ decays lead preferentially to
    final states involving $\tau$ leptons.
    Abundant multi-tau signatures constitute a severe background to
    all channels involving SUSY particles,
    in particular to the decays of interest $H,A \to \stau_i\stau_j$.
\end{itemize}

The strategy to determine the $H,A$ decay modes and branching ratios is the tagging of
one Higgs particle by its decay into a pair of $b\bar b$ jets and the analysis of the
recoiling system:
\begin{eqnarray}
  e^+ e^-& \to & H A  \ \to \ b \bar b \,X \ .
  \label{haproduction}
\end{eqnarray}
The decay modes and event topologies under investigation are
\begin{eqnarray}
  X_{\stau_i\stau_j} & = & \stau_1\stau_1 + \stau_1\stau_2 + \stau_2\stau_2
                \ \to \ \tau^+ \tau^- \,\Eslash \ ,
                \label{Xstaustau} \\
  X_{\nt_i\nt_2} & = &  \nt_1\nt_2 + \nt_2\nt_2
                \ \to \ \tau^+ \tau^- \,\Eslash  \ ,
                \label{Xni2}  \\
  X_{\cx^+_1\cx^-_1} & = &  \cx^+_1\cx^-_1
                \ \to \ \tau^+ \tau^- \,\Eslash \ ,
                \label{Xch11}
\end{eqnarray}
and the reference decay modes are
\begin{eqnarray}
  X_{\tau\tau} & = & \tau^+ \tau^-   \ ,
                \label{Xtautau} \\
  X_{b\bar b}  & = & b\bar b \to { jet \, jet} \ .
                \label{Xbb}
\end{eqnarray}
The particle content of the three supersymmetric final states
(\ref{Xstaustau}) -- (\ref{Xch11}) is identical. In order to
distinguish these channels it will be assumed that the masses of
the primary and of all the secondary SUSY particles are known well
enough so that the resulting decay topologies and $\tau^\pm$
spectra can be reliably modeled and simulated. This knowledge is
important in order to determine the branching ratios of the
various decay modes from their relative contributions to the
`observable' data distributions. This assumption is quite natural
as the measurement of the $A$ parameters is certainly a
second-generation task. Details of the event generation are
presented in the appendix~\ref{eventgeneration}.

\begin{figure} [htb] \centering
     \epsfig{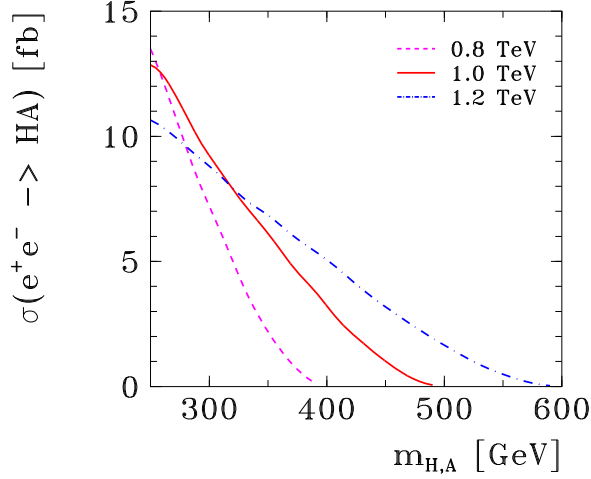}
\caption{Cross section for
  $\ee \to H A$ production as a function of the common $H,A$ mass
  at $\sqrt{s}=0.8\;\TeV, 1.0\;\TeV$ and $1.2 \;\TeV$.
  The curves include $e^\pm$ beam polarizations of
  $\cP_{e^-} = \pm 0.9$ and $\cP_{e^+} = \mp 0.6$, as well as QED radiation
  and beamstrahlung effects}
\label{sigmaha}
\end{figure}

The cross sections for $HA$ pair production \cite{Ohm} assuming common
scalar and pseudoscalar Higgs masses are shown in \fig{sigmaha}
for the three center of mass energies
$\sqrt{s} = 0.8\;\TeV$,  $1.0\;\TeV$ and $1.2\,\TeV$.
The remaining parameters are taken from the reference point
SPS1a$'$ for illustration.
The present study is representative and based on 10,000 $HA$ events which, for
scenario SPS1a$'$, may be accumulated with a cross section of 
$1.8~\fb$ at $1~\TeV$ or $3.9~\fb$ at $1.2~\TeV$, respectively.
The results may be easily scaled to lower statistics event samples without
losing their significance.

\section{Experimental analysis}

In this section the analyses of reaction (\ref{haproduction}) with
the Higgs decay modes (\ref{Xstaustau}) - (\ref{Xbb}) will be
described in detail. As mentioned above the channels involving
supersymmetric particles, $X_{\stau_i\stau_j}$, $X_{\nt_i\nt_2}$
and $X_{\cx^+_1\cx^-_1}$, lead to the same final state and the
topologies do not allow their separation on an event-by-event
basis. Rather a statistical analysis will be applied to determine
their branching fractions. The decays into Standard Model
particles, $X_{\tau\tau}$ and $X_{b\bar b}$, can be efficiently
isolated and will be used for normalization. The results will be
given in terms of combined branching ratios, defined as
$\cB_{b\bar b\,X} = \cB(H\to b\bar b)\,\cB(A\to X)
                 + \cB(A\to b\bar b)\,\cB(H\to X)$.

It should be noted that the results for the background Higgs decays
to charginos and neutralinos can probably be predicted at the time of the stau
analyses.
They depend only on parameters which can be measured in
the chargino/neutralino sector itself at earlier times. This way the
experimental results of the Higgs decays to charginos and neutralinos
can be compared with theoretical predictions.

\subsection{Signal channel {\boldmath
    $e^+ e^-\to H A \to b \bar b \; \tau^+\tau^-\Eslash$} }

The topology is characteristic for all Higgs decays into
supersymmetric particles. The criteria listed in
\tab{tab:selectioncriteria} are chosen in order to optimize the
acceptance for $HA\to b\bar b\;\stau_i\stau_j\to b\bar
b\;\tau^+\tau^-\Eslash$  \ decays.
\begin{table} \centering
\begin{tabular}{|l l|l|}
\hline
\multicolumn{2}{|l|}{Selection criteria} & Constraint \\
\hline
\hline
  { 1}  & two identified $b$ jets &                           \\
  { 2}  & $b$ jet energy      & $ 100\;\GeV < E_b < 400\;\GeV$                     \\
  { 3}  & $bb$ invariant mass & $ m_{H,A}-30\;\GeV < m_{bb} < m_{H,A}+30\;\GeV$    \\
        & recoil mass against $bb$ & $ m_{H,A}-30\;\GeV <m_{recoil} < m_{H,A}+90\;\GeV$ \\
\hline
  { 4}  & two oppositely charged $\tau$ candidates   &               \\
  { 5}  & visible $\tau$ energy       & $ 2.5\;\GeV < E_\tau < 200\;\GeV$     \\
        & visible $\tau\tau$ energy   & $E_{\tau\tau} < 250\;\GeV$             \\
\hline
  { 6}  & missing energy      & $ 250\;\GeV < \Eslash \ < 550\;\GeV$               \\
\hline
  { 7}  & acollinearity angle in Higgs rest frame & $ \xi^*_{\tau\tau} > 10^\circ$ \\
\hline
\end{tabular}
\caption{Event selection criteria for the signal reaction
  $HA\to b\bar b\;\stau_i\stau_j\to b\bar b\;\tau^+\tau^-\Eslash$ }
\label{tab:selectioncriteria}
\end{table}

The criteria (1) -- (3) provide a very efficient selection of
$HA \to b \bar b X$ events by tagging one Higgs particle via its resonant decay into a
pair of $b$ quark jets
[see discussion in appendix~\ref{eventgeneration} and \fig{hamass}].
The good energy resolution allows the reliable transformation
into the rest frame of the recoil system $X$ which is identified as the second
Higgs particle.

\begin{figure}
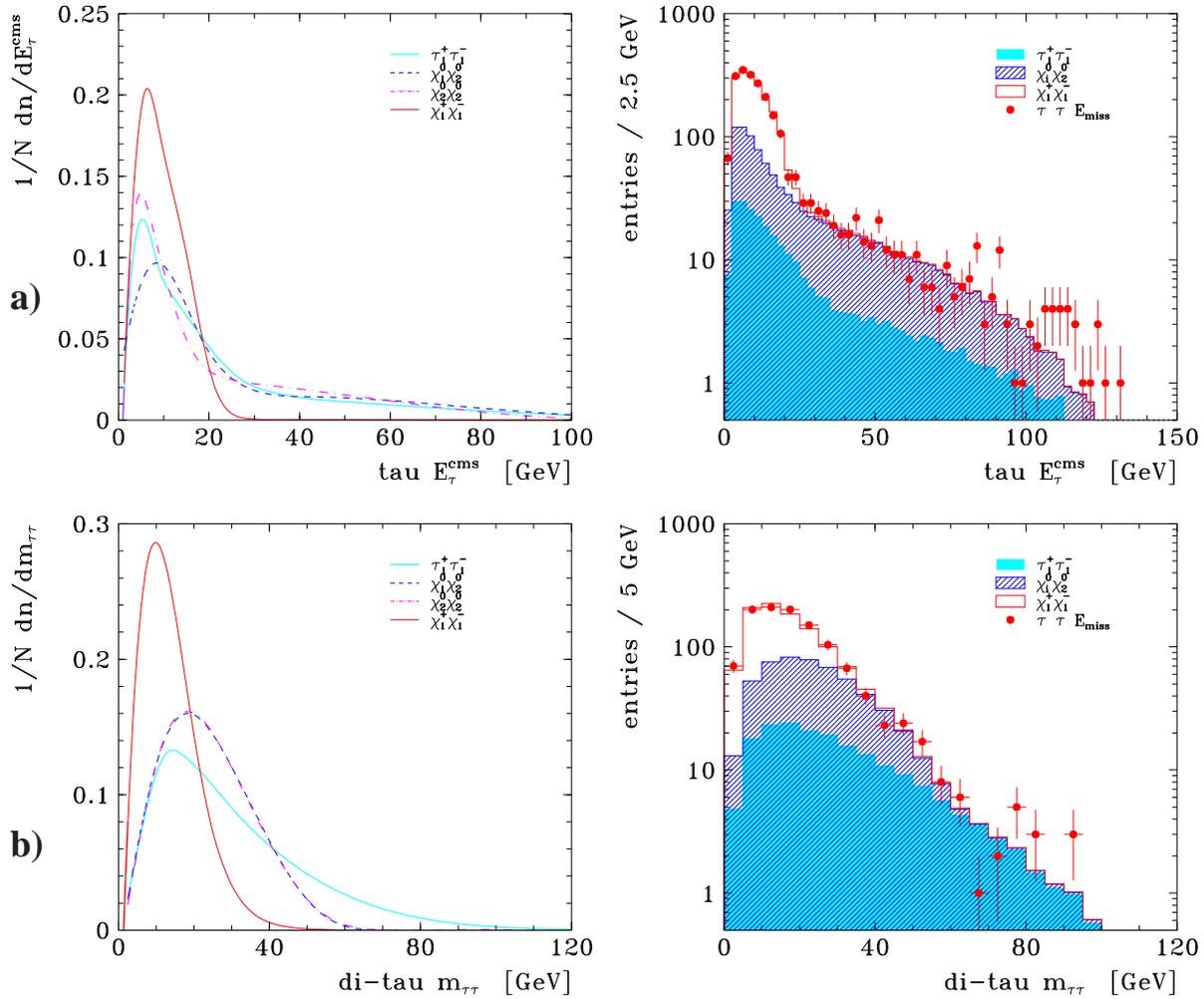

  {   \epsfig{file=habr_fetcm.eps,angle=90,%
          width=.5\textwidth}
      \epsfig{file=habr_etcml.eps,angle=90,%
          width=.5\textwidth} }
  {    \epsfig{file=habr_fmtt.eps,angle=90,%
          width=.5\textwidth}
       \epsfig{file=habr_mttl.eps,angle=90,%
          width=.5\textwidth} }
\begin{picture}(100,0)(0,0)
  \put(0, 97){\large\bf a) }
  \put(0, 24){\large\bf b) }
\end{picture}
\caption{Spectra from
  $\ee\to HA \to b \bar b\; \tau^+\tau^-\Eslash$ \ decays of
  a) visible tau energy $E^{*}_{\tau}$ in the $H,A$ rest frame;
  b) di-tau mass $m_{\tau\tau}$.
  Left: normalized distributions; right:
  fitted contributions of individual channels
  $\stau_i\stau_j$, $\nt_i\nt_2$ and $\cx^+_1\cx^-_1$
  to the observable signal.
  mSUGRA scenario SPS1a$'$ at $\sqrt{s}=1\;\TeV$}
\label{HA-spectra}
\end{figure}

The criteria (4) -- (6) select SUSY decays into secondary $\tau's$
plus large missing energy. The last cut (7) removes direct decays
into $\tau\tau$ pairs, which are back-to-back in the Higgs rest
frame. The properties of the various decay modes are displayed in
the left panels of \fig{HA-spectra}, where normalized
distributions of the visible tau energy and di-tau mass are shown.
It is a common feature of both spectra that the dominant
contributions come from $\cx^+_1\cx^-_1$ peaks at low values,
while the spectra from $\stau_i\stau_j$ and $\nt_i\nt_2$ extend
towards higher values. The separation between the decay modes
improves when correlations between both particles are exploited,
{\em e.g.} the invariant mass $m_{\tau\tau}$. Also notice that the
shapes of the distributions from $\nt_1\nt_2$ and $\nt_2\nt_2$ are
barely distinguishable, thus only the sum of all neutralino
decays, labeled $\nt_i\nt_2$, will be investigated.

The overall $H,A\to\stau_i\stau_j$ efficiency is $\sim 43\%$.
However, there are still large contributions from Higgs decays into
charginos ($\sim 37\%$)  and neutralinos ($\sim 23\%$),
both of which have higher combined branching
ratios, see \tab{tab:decaymodes}.

The distributions from the complete simulation
of the visible $\tau$ energy in the Higgs rest frame $E_\tau^*$
and the di-tau mass $m_{\tau\tau}$
are shown in the right panels of \fig{HA-spectra}.
The contributions from the individual decay modes,
$X_{\stau_i\stau_j}$, $X_{\nt_i\nt_2}$ and $X_{\cx^+_1\cx^-_1}$
of eqs.~(\ref{Xstaustau}) - (\ref{Xch11}),
are summed up and fitted to reproduce the data of \fig{HA-spectra}.
The analyses of the observables $E^*_{\tau}$ 
and $m_{\tau\tau}$ emphasize different characteristics but lead to
consistent results. In all fits (including observables not shown)
the chargino contribution can be determined in a stable manner
whereas the stau and neutralino parts are strongly correlated.

The fit results are displayed in the spectra of \fig{HA-spectra}.
The relative rates, acceptances and the extracted combined
branching ratios $\cB_{b \bar b X}$ are summarized in
\tab{tab:results}.

\begin{table}[htb] \centering
\begin{tabular}{|l| c c| c|}
\hline
  $\ee\to H A \to b \bar b X$
              & $f_{b \bar b X}^{\rm fit}$
              & $\epsilon_{b \bar b X}$
              & $\cB_{b \bar b X}$ \\
\hline
\hline
  $H A \to b \bar b\,\stau_i\stau_j$
              & $0.186 \pm 0.041$  & $0.428$
              & $0.049 \pm 0.011$    \\
  $\phantom{H A \to }\; b \bar b\,\nt_i\nt_2$
              & $0.292 \pm 0.052$  & $0.228$
              & $0.135 \pm 0.024$    \\
  $\phantom{H A \to }\; b \bar b\,\cx^+_1\cx^-_1$
              & $0.516 \pm 0.036$  & $0.372$
              & $0.146 \pm 0.010$    \\
\hline
  $H A \to b \bar b\,\tau \tau$
              & $          $       & $0.515$
              & $0.075 \pm 0.004$    \\
\hline
  $H A \to b \bar b\,b \bar b$
              & $          $       & $0.630 $
              & $0.345 \pm 0.007$    \\
\hline
\end{tabular}
\caption{Expected accuracies on the determination of Higgs decays
  $H A\to b \bar b X$.
  Listed are the analyzed event samples,
  the fitted contributions of decay modes $f_{b \bar b X}^{\rm fit}$ ,
  the detection efficiencies $\epsilon_{b \bar b X}$,
  and the combined branching ratios $\cB_{b \bar b X}$.
  The results are based on 10,000 $HA$ decays in the SPS~1a$'$ scenario}
\label{tab:results}
\end{table}

\subsection{Reference channels {\boldmath
    $e^+ e^-\to H A \to b \bar b \; \tau \tau\,$ and
      $\,H A \to  b \bar b \;  b \bar b$} }

\paragraph{\boldmath $e^+ e^-\to HA \to b\bar b\;\tau^+\tau^- \,$}

The selection of $HA \to b\bar b\;\tau^+\tau^-$ decays is
complementary to the analysis of the previous SUSY decays.
The basic criteria (1) -- (4) of \tab{tab:selectioncriteria}
for $bb$ and $\tau\tau$ identification are applied.
However, the $\tau$ energy spectra are harder and both $\tau's$ are emitted
back-to-back in the Higgs rest frame, leading to the following cuts:
(5) visible $\tau$ energy $ 5 < E_\tau < 400\;\GeV$,
    $\tau\tau$ energy  $E_{\tau\tau} < 500\;\GeV$;
(6) no missing energy requirement;
(7) acollinearity angle in the Higgs rest frame
    $\xi^*_{\tau\tau} < 10^\circ$.

\begin{figure}[htb] \centering
  \epsfig{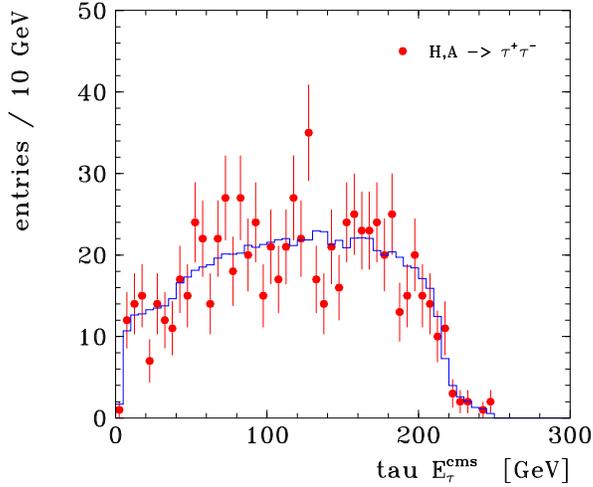}
  \caption{Spectrum of the visible
    $\tau$ energy $E^{*}_{\tau}$ in the $H,A$ rest frame from
    $\ee\to HA \to b \bar b \;\tau^+\tau^-$ \ decays.
    mSUGRA scenario SPS1a$'$ at $\sqrt{s}=1\;\TeV$}
  \label{tautau-spectra}
\end{figure}

The reconstructed spectrum of the visible $\tau$ energy $E_\tau^*$
in the Higgs rest frame,
shown in \fig{tautau-spectra}, is fairly flat and extends up to the energy of
the primary, undecayed $\tau$ lepton.
The overall detection efficiency is high,
see \tab{tab:results}.
A combined branching ratio
of $\cB_{b \bar b\, \tau\tau} = 0.075 \pm 0.004$
can be obtained,
where only statistical uncertainties are given.
The analysis may be further
improved by an overconstrained kinematical fit.
Exploiting energy-momentum conservation and
approximating the $\tau$ directions by the directions of the
decay products and treating the $\tau$ energies as free parameters,
allows one to construct 2 constraints (2-C fit),
see~\cite{desch}.

\begin{figure}[htb] \centering
  \epsfig{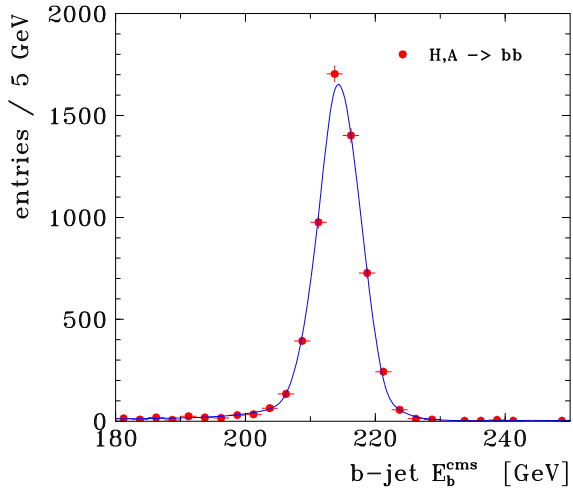}
  \caption{Spectrum of
    $b$-jet energy $E_{b}^*$ in the $H,A$ rest frame from
    $\ee\to HA \to b \bar b \; b \bar b$ \ decays.
    mSUGRA scenario SPS1a$'$ at $\sqrt{s}=1\;\TeV$}
  \label{bb-spectra}
\end{figure}

\paragraph{\boldmath
    $e^+ e^-\to H A \to b \bar b \;  b \bar b\,$}
The selection of $HA\to b \bar b \, b \bar b$ events is straightforward
by applying the same  criteria (1) -- (3) of
\tab{tab:selectioncriteria} to another pair of $bb$ jets.
The four jets are then combined such as to construct two $bb$ systems with
invariant masses closest to each other,
$m_{bb}^{(1)} \simeq m_{bb}^{(2)}$.
Again, the selection efficiency is high.
The energy distribution of the $b$ jets in the Higgs rest frame,
displayed in \fig{bb-spectra}, exhibits a clear signal
of a narrow peak at half the Higgs mass.
The combined branching ratio for the decay mode $X_{b\bar b}$
can be determined with a statistical accuracy of
$\cB_{b\bar b\, b\bar b} = 0.345 \pm 0.007$, for details
see \tab{tab:results}.

The measurement only provides information on the product of cross
section times branching ratio. In order to extract the decay rates
the $HA$ production cross section has to be calculated accurately,
which in turn requires a precise knowledge of the Higgs masses.
The $b$ jet energy distribution of \fig{bb-spectra} (or
equivalently  the $bb$ mass spectrum similar to \fig{hamass}a) can
be used to determine the $H,A$ masses with an accuracy of $\delta
m_{H,A} \simeq 0.15~\GeV$. This error can be reduced further by
applying kinematic fitting techniques~\cite{desch}. In fact such a
procedure allows the selection of a very clean $HA$ event sample
with low background, so that relaxing of the $b$ quark
identification criteria may be envisaged.

\section{Interpretation and conclusions}

The expected results for the combined branching ratios of
Higgs decay modes (\ref{Xstaustau}) -- (\ref{Xbb})
are summarized in \tab{tab:results}.
The double ratios are experimentally determined as
$ \cB_{b\bar b \,\stau_i\stau_j}/\cB_{b\bar b\, b\bar b}
   =  0.142 \pm 0.032$ and
$ \cB_{b\bar b \,\stau_i\stau_j}/\cB_{b\bar b\, \tau\tau}
   =  0.653 \pm 0.147 $.
Their relations to the partial decay widths can be written
as\footnote{As mentioned earlier,
  the pseudoscalar Higgs boson $A$ decays only to off-diagonal
  $\stau_1\stau_2$ pairs in
  CP-invariant theories while the scalar Higgs boson $H$ can decay to all
  combinations of stau pairs }
\begin{eqnarray}
  \frac{\cB_{b\bar b \,\stau_i\stau_j }} {\cB_{b\bar b\, b\bar b} }
  & = & \frac{\cB(H\to b\bar b)\,\cB(A\to \stau_1\stau_2)
            + \cB(A\to b\bar b)\,\cB(H\to \stau_i\stau_j)}
             { \cB(H\to b\bar b)\,\cB(A\to b\bar b)} \nn \\[.2em]
  & = & \frac{\Gamma(A\to \stau_1\stau_2)}{\Gamma(A\to b\bar b)}
      + \frac{\Gamma(H\to \stau_i\stau_j)}{\Gamma(H\to b\bar b)} \ ,
      \label{eq:rbb}
  \\[1em]
  \frac{\cB_{b\bar b\,\stau_i\stau_j}} {\cB_{b\bar b\, \tau\tau} }
  & = & \frac{\cB(H\to b\bar b)\,\cB(A\to \stau_1\stau_2)
            + \cB(A\to b\bar b)\,\cB(H\to \stau_i\stau_j)}
             {\cB(H\to b\bar b)\,\cB(A\to \tau\tau)
            + \cB(A\to b\bar b)\,\cB(H\to \tau\tau)} \phantom{xxxxxxxxxxx}
      \nn \\[.2em]
  & = & \frac{\Gamma(A\to \stau_1\stau_2)}
             {\Gamma(A\to \tau\tau) \,(1+r)}
      + \frac{\Gamma(H\to \stau_i\stau_j)}
             {\Gamma(H\to \tau\tau) \,(1+1/r)}
      \nn \\[.2em]
& = & \frac{1}{2}\,\left [ \, \frac{\Gamma(A\to \stau_1\stau_2)}
                                   {\Gamma(A\to \tau\tau)}
                            + \frac{\Gamma(H\to \stau_i\stau_j)}
                                   {\Gamma(H\to \tau\tau)} \, \right ]  \ .
      \label{eq:rtt}
 \end{eqnarray}
In the second double ratio the two terms in the denominator have been
identified, {\it i.e.}
\begin{eqnarray}
  r & = & \frac{\Gamma(A\to b\bar b)\;\Gamma(H\to\tau\tau)}
                {\Gamma(H\to b\bar b)\;\Gamma(A\to\tau\tau)}
    \ = \ 1 \ ,
\end{eqnarray}
which is expected to hold with high accuracy.

These double ratios are proportional to
$(A_\tau + \mu \cot\beta)^2 \simeq A^2_\tau$
in the decoupling limit for large $A_\tau$ and large $\tan\beta$ when
$LR$ decays dominate over the diagonal $LL$ and $RR$ decays and mixing can be
neglected. For the parameters chosen in this study, however, we must
include corrections from $LR$ mixing of the particles and the diagonal $LL$ and $RR$
decays.

\begin{figure}[htb] 
  { \epsfig{file=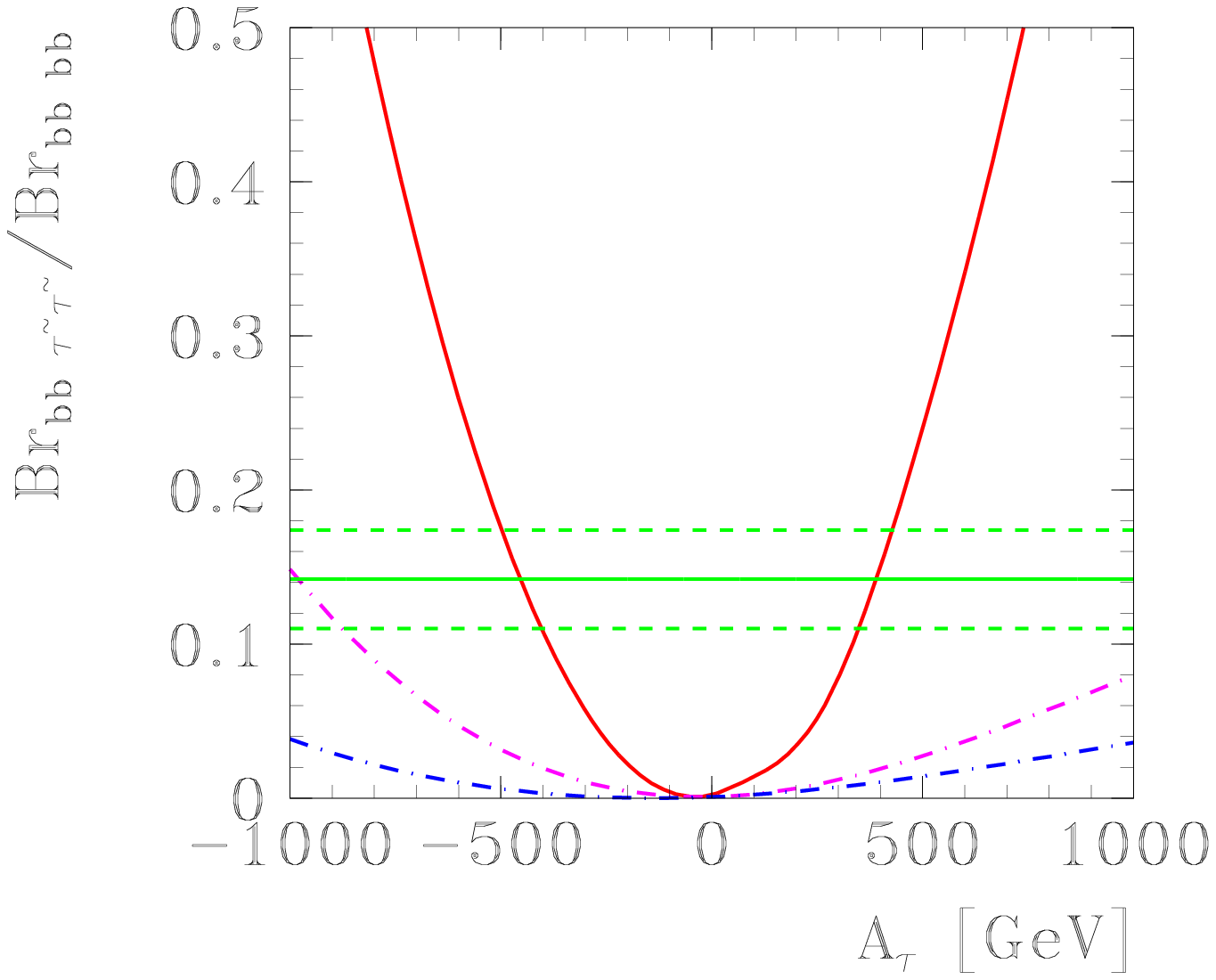,angle=90,width=.5\textwidth}
    \epsfig{file=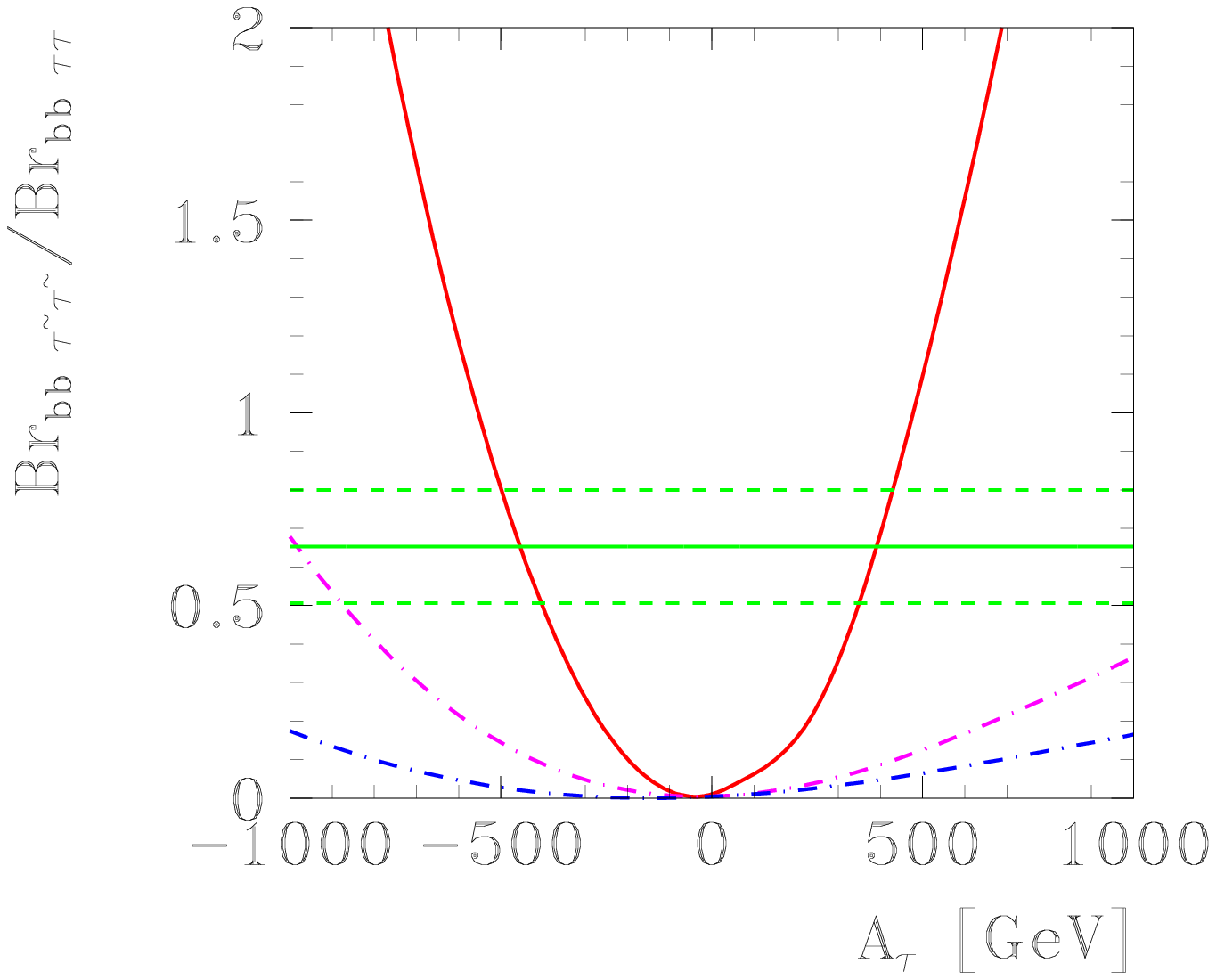,angle=90,width=.5\textwidth} }
  \caption{Double ratios of the combined branching ratio
    ${\cal B}_{HA \to b \bar b \, \stau_i\stau_j}$ normalized to
    ${\cal B}_{HA \to b \bar b \, b \bar b}$ (left) and
    ${\cal B}_{HA \to b \bar b \, \tau\tau}$ (right)
    as a function of $A_\tau$.
    The lower curves show the contributions from diagonal pairs
    $\stau_2\stau_2$ (blue) and $\stau_1\stau_1$ (magenta).
    The horizontal (green) lines indicate the expected experimental accuracy
    based on 10,000 $HA$ decays in scenario SPS1a$'$ 
  }
  \label{stau_ratios}
\end{figure}

The pseudoscalar Higgs boson $A$ couples to off-diagonal $ \stau_1 \stau_2$ pairs
with the same amplitude as to  $ \stau_L \stau_R$ pairs so that no mixing
corrections need be applied. In contrast, the coupling of the scalar Higgs boson $H$
to off-diagonal stau pairs is modified by the mixing parameter $\cos 2 \theta_{\stau}$
and, moreover, $H$ decays include also contributions from the genuine diagonal
couplings $ \stau_L \stau_L$
and $ \stau_R \stau_R$ of the order $m_\tau / A_\tau$
and $m_Z / (A_\tau \tan\beta)$ with respect to the leading off-diagonal
amplitudes. Since the mixing parameter
\begin{eqnarray}
  \sin 2\theta_{\stau} & = & \frac{2\,m_\tau}{m^2_{\stau_1} - m^2_{\stau_2} }\,
                \left ( A_\tau - \mu\,\tan\beta \right )
\end{eqnarray}
involves $A_\tau$ itself, the dependence of the decay amplitudes
on $A_\tau$ is modified and not
linear anymore. As a result, the binomial character of the partial decay widths
in $A_\tau$ is changed asymmetrically.

The dependence of the double ratios of the decay widths, {\it cf}
eqs.~(\ref{eq:rbb}), (\ref{eq:rtt}) and (\ref{eq:ratiostautau}),
on $A_\tau$ including the subleading $LR$ mixing effects are
calculated using {\sc FeynHiggs}~\cite{feynhiggs} and are
displayed in \fig{stau_ratios} together with the expected
experimental accuracies. There are two possible solutions for
$A_\tau$ at $-450\;\GeV$ and $+350\;\GeV$, which, however, can be
distinguished experimentally because they correspond to different $\stau$
mixing configurations.  
The mixing parameter $\sin 2\theta_{\stau}$
differs by $\sim~20\,\%$ for the two solutions. Since the mixing
can be obtained with an accuracy of a few percent from
measurements of the $\stau_1$ mass and the $\stau_1 \stau_1$
production cross section at $e^+e^-$ colliders (see
\cite{martyn,boos} for scenarios with similar parameters), this
additional information is sufficient to single out the
negative $A_\tau$ solution. 

From the simulation of heavy Higgs decays into supersymmetric and SM particles
one obtains for the trilinear coupling
\begin{eqnarray}
  A_\tau & = & -450 \pm 50 ~\GeV \nn
  \label{eq:result}
\end{eqnarray}
as an {\it ab-initio} determination of this soft SUSY breaking
parameter for an event sample of 10,000 $HA$ Higgs pairs.

Uncertainties in the theoretical predictions and parameters,
eqs~(\ref{eq:ratiostautau}) --  (\ref{eq:ratiobb})
and their analogues for diagonal decays,
are expected to be negligible at the level of achievable experimental accuracies.
Theoretical calculations are under control at the per-cent level
when all the one-loop corrections in the $\stau/\tau$ and Higgs sectors
are included \cite{RC}. It is interesting to note that
the parameter $\tan\beta$ can be controlled internally within
the same analysis of $HA\to b \bar b \, b \bar b$ decays.
The $\tan\beta$ dependence of the combined branching ratios can be expressed 
as $\cB_{b\bar b b\bar b}= 1/[(1 +c_A/\tan^2\beta)(1+c_H/\tan^2\beta)]$
with coefficients $c_A\simeq 100$ and $c_H\simeq 50$ for $A$ and $H$
decays, respectively, in the reference point considered.
From the measurement quoted in \tab{tab:results}
one expects a precision of $\delta\tan\beta\simeq 0.15$.
The parameter $\mu$ can be measured in chargino production within a few
per-cent.
Both uncertainties result in a shift of the trilinear coupling of at most
$\delta A_\tau \lesssim 1\,\GeV$, far below the anticipated experimental error.
These estimates are confirmed by a combined analysis of SUSY parameters
based on measurements of many SUSY production processes at the ILC and
LHC~\cite{bechtle}.

The direct determination of the trilinear coupling analyzed in the present report
may be compared with other methods which
make use of higher order corrections affected by the parameter $A_\tau$.
A global analysis by means of Fittino~\cite{bechtle} provides a combination
of $X_\tau = A_\tau - \mu\tan\beta = -4450 \pm 30~\GeV$, together with
$\tan\beta  =  10.0 \pm 0.1$ and
$\mu  =  400.4 \pm 1.3~\GeV$.
However, application of this indirect method is possible {\it a priori} only
in scenarios
in which the degrees of freedom are specified {\it in toto} when
the virtual loop corrections are included and if all theoretical uncertainties
are under proper control. In contrast, the proposal described in
the present paper is a robust leading order analysis.

\section{Summary}

While the trilinear stop-Higgs coupling $A_t$ can be measured
fairly easily by evaluating the stop masses and the mixing angle,
this task is much more demanding for the trilinear coupling
$A_\tau$ in the stau sector since these couplings come with
the masses of the quarks and leptons. Nevertheless, we have
demonstrated in this report that the measurement of $A_\tau$
is possible in scalar and pseudoscalar Higgs boson $H,A$ decays.
Large luminosities at the $e^+e^-$ linear collider ILC would
be required, however, to achieve an accuracy of about 10\%. Though
the measurement is difficult, this direct determination based
on tree-level processes is necessary before the determination
through indirect effects
based on quantum corrections can be trusted with high confidence.

After the stop trilinear coupling $A_t$ will be determined,
the measurement of at least one
additional trilinear parameter is required to investigate
universality properties of these parameters, for instance,
as implemented in
minimal supergravity. The $A_\tau$ measurements are therefore
important ingredients for reconstructing the
underlying physics scenario \cite{Blair}.

\paragraph{Acknowledgement}
We are grateful to S. Heinemeyer for helpful communications
on the FeynHiggs program.
The work of SYC was supported in part by the Korea Research
Foundation Grant (KRF--2002--041--C00081) and in part by KOSEF
through CHEP at Kyungpook National University. PMZ thanks the
German Science Foundation DFG and Stanford University
for partial financial support during an extended visit of 
SLAC where part of this work was carried out.

\appendix
\section{Event generation}
  \label{eventgeneration}

Events are generated with the program \pythia~6.3~\cite{pythia}
which includes initial and final state QED radiation as well as
beamstrahlung \`a la {\sc Circe}~\cite{circe}.
The decays of $\tau$ leptons are treated by {\sc Tauola}~\cite{tauola}.
The  detector simulation is based on the detector proposed
in the {\sc Tesla tdr}~\cite{tdr}
and implemented in the Monte Carlo program {\sc Simdet}~4.02~\cite{simdet}.
The main detector features are excellent particle identification and
measurement of charged and neutral particles
for a polar angle acceptance $\theta \; (\pi-\theta) > 125~\mrad$.

\begin{figure}[htb]
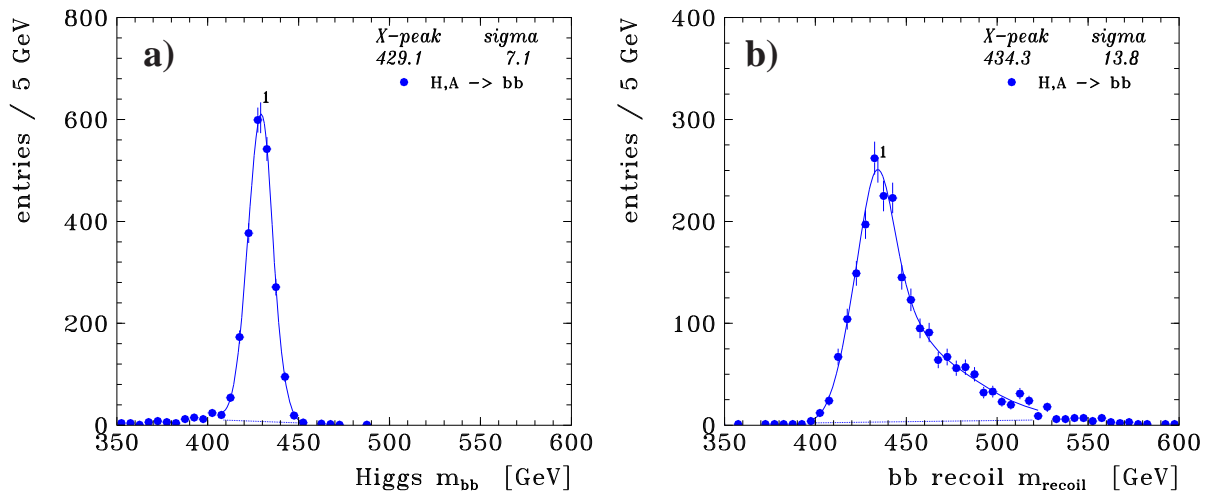

  \epsfig{file=habr_mbb.eps,angle=90,%
    width=.5\textwidth}
  \epsfig{file=habr_mbbrecoil.eps,angle=90,%
    width=.5\textwidth}
  \begin{picture}(100,0)(0,0)
    \put( 18, 62){\large\bf a) }
    \put( 98, 62){\large\bf b) }
  \end{picture}
  \caption{Spectra from
    $\ee \to H A \to b \bar b \; X$ \ decays of
    a) $bb$ di-jet mass $m_{b b}$;
    b) recoil mass  $m_{recoil}$ against $b b$ system.
    mSUGRA scenario SPS1a$'$ at $\sqrt{s}=1\;\TeV$}
  \label{hamass}
\end{figure}

In the analysis the reconstructed $b$ jets and $\tau$ candidates are
required to be within the acceptance of $|\cos\theta|<0.95$,
while Higgs boson are produced centrally $\sim\sin^2\theta$.
The identification of $HA\to b\bar b\,X$ events is provided by the good jet
energy flow measurement with a
resolution of $\sigma/E=0.3/\sqrt{E(\GeV)}$.
This is illustrated in \fig{hamass}, where
the di-jet mass and the mass recoiling against the $b\bar b$ system
are shown.
Both distributions are fairly narrow and peak at the Higgs masses.
The recoil mass spectrum is slightly wider and extends towards large
values due to radiative effects.

For the identification and reconstruction
of $\tau$ candidates, a narrow jet with invariant mass $m_\tau<2.5\;\GeV$
is required which contains
one charged particle plus possibly additional photons
or three charged particles.
In general the leptonic 3-body decays
$\tau\to e\nu_e\nu_\tau$ (17.8\%),
$\tau\to\mu\nu_\mu\nu_\tau$ (17.4\%)
are less sensitive to the primary $\tau$ energy than
the hadronic decays
$\tau\to\pi\nu_\tau$ (11.1\%),
$\tau 
     \to\pi^\pm\pi^0\nu_\tau$ (25.4\%) and
$\tau 
     \to \pi^\pm \pi^+\pi^-\nu_\tau +
  \pi^\pm\pi^0\pi^0 \nu_\tau$ (19.4\%).
All decay modes are used in the analysis, except of $ee$ and $\mu\mu$ pairs.

Since the decay rates of interest are of the order of a few
percent, to be further degraded by efficiency losses, the case
study is based on a high statistics sample assuming a production
rate of $N_{HA} = \sigma_{HA}\cdot\cL = 10,000$ events. The total
cross section of $HA$ production amounts to $\sigma_{HA} =
1.8\;\fb$ at $\sqrt{s} = 1\;\TeV$, including $e^\pm$ beam
polarization, QED radiation and beamstrahlung, see \fig{sigmaha}.
The results of the present study may be easily transferred to any
other energy or reference point once the parameters are specified.

The characteristic event signatures $HA \to b \bar b \, \tau\tau\,\Eslash$,
{\em i.e.} two energetic $b$ jets forming a high mass resonant state plus two
$\tau$ leptons plus (possibly large) missing energy, are very clean.
Any background from QCD processes $q \bar q (g)$,
$WW$ or $ZZ$ production is estimated to be
small and is therefore neglected.

%


\begin{thebibliography}{99}

\bibitem{Djou}
        A.~Djouadi, Review {\it The Higgs Bosons in the Minimal
            Supersymmetric Model},
        Report LPT-ORSAY-05-18, arXiv:hep-ph/0503173.

\bibitem{bartl}
        A.~Bartl, H.~Eberl, S.~Kraml, W.~Majerotto and W.~Porod,
        Eur.\ Phys.\ J.\ direct {\bf C2} (2000) 6.

\bibitem{boos}
  E.~Boos, H.-U.~Martyn, G.~Moortgat-Pick, M.~Sachwitz, A.~Sherstnev and P.~M.~Zerwas,
  Eur.\ Phys.\ J.\ C {\bf 30} (2003) 395.

\bibitem{spa}
        {\it Supersymmetry Parameter Analysis: SPA Convention
          and Project},
        http://spa.desy.de/spa; \\
        B. C.~Allanach {\it et al.}, Eur. Phys. J. {\bf C25} (2002) 113.

\bibitem{feynhiggs}
        S.~Heinemeyer, W.~Hollik and G.~Weiglein,
        Comput.\ Phys.\ Commun.\  {\bf 124} (2000) 76.

\bibitem{Ohm}
        A.~Djouadi, J.~Kalinowski, P.~Ohmann and P. M.~Zerwas,
        Z. Phys. {\bf C74} (1997) 93.

\bibitem{tdr} {\sc Tesla} Technical Design Report, DESY 2001-011,
        Part III: {\em Physics at an $e^+e^-$ Linear Collider}
        [arXiv:hep-ph/0106315],
        Part IV: {\em A Detector for TESLA}.

\bibitem{desch}
        K.~Desch, T.~Klimkovich, T.~Kuhl and A.~Raspereza,
        Linear Collider note LC-PHSM-2004-006,
        arXiv:hep-ph/0406229.

\bibitem{martyn}
  H.-U.~Martyn,
  contribution to {\it 3rd Workshop of ECFA/DESY LC Study, 2002, Prague},
  LC-PHSM-2003-071,
  arXiv:hep-ph/0406123.

\bibitem{RC}
  J.~Guasch, W.~Hollik and J.~Sola,
  JHEP {\bf 0210} (2002) 040; \\
  S. Heinemeyer, W. Hollik, J. Rosiek and G. Weiglein,
  Eur. Phys. J. {\bf C19} (2001) 535.

\bibitem{bechtle}
  P.~Bechtle, K.~Desch and P.~Wienemann,
  Proceedings {\it International Linear Collider Workshop 2005, Stanford, USA},
  arXiv:hep-ph/0506244.

\bibitem{Blair}
  G.A.~Blair, W.~Porod and P.M.~Zerwas, Phys. Rev. {\bf D63} (2001) 017703 and
  Eur. Phys. J. {\bf C27} (2003) 263.

\bibitem{pythia} 
        T.~Sj\"ostrand {\it et al.},
        Comput. Phys. Commun. {\bf 135} (2001) 238.

\bibitem{circe} 
        T.~Ohl,
        Comput. Phys. Commun. {\bf 101} (1997) 269.

\bibitem{tauola} 
        S.~Jadach {\it et al.},
        Comput. Phys. Commun. {\bf 76} (1993) 361.

\bibitem{simdet} 
        M.~Pohl and J.~Schreiber,
        DESY~02-061, arXiv:hep-ex/0206009.

\end{thebibliography}
\end{document}